\documentclass[prx,aps,twocolumn,superscriptaddress]{revtex4}

\usepackage{dcolumn}
\usepackage{bm}
\usepackage{amssymb}
\usepackage{color}
\usepackage[pdftex]{graphicx}

\usepackage{amsmath}
\usepackage[linesnumbered,ruled]{algorithm2e}

\begin{document}


\title{Semiclassical  dynamics of spin density waves}

\author{Gia-Wei Chern}
\affiliation{Department of Physics, University of Virginia, Charlottesville, VA 22904, USA}

\author{Kipton Barros}

\affiliation{Theoretical Division, Los Alamos National Laboratory, Los Alamos, New Mexico 87545, USA}

\author{Zhentao Wang}

\affiliation{Department of Physics and Astronomy, The University of Tennessee, Knoxville, TN 37996, USA}

\author{Hidemaro Suwa}

\affiliation{Department of Physics and Astronomy, The University of Tennessee, Knoxville, TN 37996, USA}

\author{Cristian D. Batista}

\affiliation{Department of Physics and Astronomy, The University of Tennessee, Knoxville, TN 37996, USA}
\affiliation{Quantum Condensed Matter Division and Shull-Wollan Center, Oak Ridge National Laboratory, Oak Ridge, Tennessee 37831, USA}

\date{\today}

\begin{abstract}
We present a theoretical framework for equilibrium and nonequilibrium dynamical simulation of quantum states with spin-density-wave (SDW) order. Within a semiclassical adiabatic approximation that retains electron degrees of freedom, we demonstrate that the SDW order parameter obeys a generalized Landau-Lifshitz equation. With the aid of an enhanced kernel polynomial method, our linear-scaling quantum Landau-Lifshitz dynamics (QLLD) method enables dynamical SDW simulations with $N \simeq 10^5$ lattice sites. Our real-space formulation can be used to compute dynamical responses, such as dynamical structure factor, of complex and even inhomogeneous  SDW configurations at zero or finite temperatures. Applying the QLLD to study the relaxation of a noncoplanar topological SDW under the excitation of a short pulse, we further demonstrate the crucial role of spatial correlations and fluctuations in the SDW dynamics.
\end{abstract}

\maketitle

Quantum states with unusual broken symmetries have long fascinated physicists because of their fundamental importance and potential technological applications. Of particular interest is the regular spatial modulation of electron spin
known as spin-density wave~(SDW) state~\cite{fawcett88,gruner94}. SDWs are ubiquitous in strongly correlated systems and play a crucial role in several intriguing many-body phenomena. For example, the SDW state is proximate to the superconducting phase in several unconventional superconductors, including cuprates and iron pnictides. Indeed, non-Fermi liquid behavior is usually observed in the vicinity of a SDW phase transition~\cite{ritz13}. Moreover, conduction electrons propagating in a noncoplanar spin texture acquire a nontrivial Berry phase and exhibit unusual transport and topological properties~\cite{roessler06,nagaosa10}.  Consequently, metallic SDW with complex spin structures, such as spirals or skyrmion crystals, offers a novel route to control the charge degrees of freedom through manipulation of spins and vice versa~\cite{nagaosa13}.  

While analytical techniques have yielded much insight about itinerant magnetism and SDW states~\cite{moryia85,brando16}, numerical methods continue to provide valuable benchmarks and shed light on controversial issues. Among the various numerical tools~\cite{simon_collab}, quantum Monte Carlo (QMC) simulations provide numerically exact solutions to strongly correlated models~\cite{blankenbecler81,hirsch85,white89}. However, one severe restriction of most QMC methods is the infamous sign-problem. Powerful alternative  approaches that avoid the sign-problem include dynamical mean-field theory (DMFT)~\cite{georges96,kotliar06} and density-matrix renormalization group (DMRG)~\cite{white92,schollwoeck04}.  Significant developments have also been made in their nonequilibrium extension such as time-dependent (TD) DMFT~\cite{freericks06,aoki14} and TD-DMRG~\cite{white04,daley04}. Both methods, however, are still very limited in their treatment of complex mesoscopic structures.

In this paper, we present a different numerical approach to SDW dynamics, emphasizing the ability to simulate large-scale lattices and complex SDW orders that often occur in highly frustrated systems. Our starting point is a semi-classical treatment of Hubbard-like models, which  {\it neglects quantum fluctuations, but retains the spatial fluctuations of the SDW field. In a way, this approach is the complement of DMFT, which includes quantum fluctuations at the expense of neglecting spatial correlations.} A systematic approach is then developed to reintroduce quantum dynamics to the SDW order parameter. We show that in the leading adiabatic approximation, the SDW dynamics is described by a generalized Landau-Lifshitz (LL) equation in which the effective forces acting on the spins are generated from itinerant electrons. Our numerical scheme can be viewed as a quantum LL dynamics (QLLD), in which the electronic degrees of freedom are integrated out at each time step. By supplementing the LL equation with Ginzburg-Landau type relaxation and stochastic terms,  our QLLD method can be used to simulate SDW dynamics both near and far-from equilibrium.

\section{Spin-Fermion Hamiltonian for equilibrium SDW phases}
\label{sec:GL-dyn}

We consider the one-band Hubbard model with an on-site repulsion $U > 0$,
\begin{eqnarray}
	\label{eq:H_hubbard}
	\mathcal{H} = - \sum_{ij, \alpha} t_{ij} \,c^\dagger_{i, \alpha} c^{\;}_{j, \alpha} + U \sum_i n_{i,\uparrow} n_{i\,\downarrow}.
\end{eqnarray}
After performing the Hubbard-Stratonovich (HS) transformation~\cite{schulz90,mukherjee14}, we obtain the following spin-fermion Hamiltonian
\begin{eqnarray}
	\label{eq:H_SDW}
	\mathcal{H}_{\rm SDW} = - \sum_{ij, \alpha} t_{ij} \,c^\dagger_{i, \alpha} c^{\;}_{j, \alpha} - 2 U \sum_i \mathbf m_i \cdot \mathbf s_i + U \sum_i |\mathbf m_i|^2, \,\,
\end{eqnarray}
where $\mathbf s_i = \frac{1}{2} c^\dagger_{i,\alpha} \bm\sigma_{\alpha\beta} c^{\;}_{i, \beta}$ is the spin operator of conduction electrons and $\bm\sigma$ is a vector of the Pauli matrices. The local HS or auxiliary field $\mathbf m_i$ is a classical O(3) vector in $\mathbb{R}^3$.
Here we have set $\hbar = 1$.  Importantly, since $\mathcal{H}_{\rm SDW}$ describes {\rm non-interacting} electrons coupled to a magnetic background, the fermionic degrees of freedom can be integrated out either in Monte Carlo or dynamical simulations to be described below.

This HS Hamiltonian is typically the starting point for determinant QMC (DQMC) simulations~\cite{blankenbecler81,hirsch85,white89}. In an alternative approach, one assumes static HS variables; then the above Hamiltonian resembles the so-called spin-fermion model~\cite{buhler00,alvarez02} and can be simulated with Markov chain Monte Carlo assuming classical ``spins'' $\mathbf m_i$, while electrons are treated quantum mechanically. Applying this method to the cubic-lattice Hubbard model, the obtained N\'eel temperature agrees remarkably well with those from DQMC simulations~\cite{mukherjee14}. It is worth noting that while Eq.~(\ref{eq:H_SDW}) is similar to the Hartree-Fock treatment of the Hubbard model, retaining spatial fluctuations of the local HS fields $\{\mathbf m_i\}$ in this static (in imaginary time) HS-field formalism goes beyond the usual mean-field method. For instance, {\it this approach captures the critical fluctuations, and consequently the correct universality class, of any continuous thermodynamic transition into a magnetically ordered state.}

\begin{figure}[t]
\includegraphics[width=0.95\columnwidth]{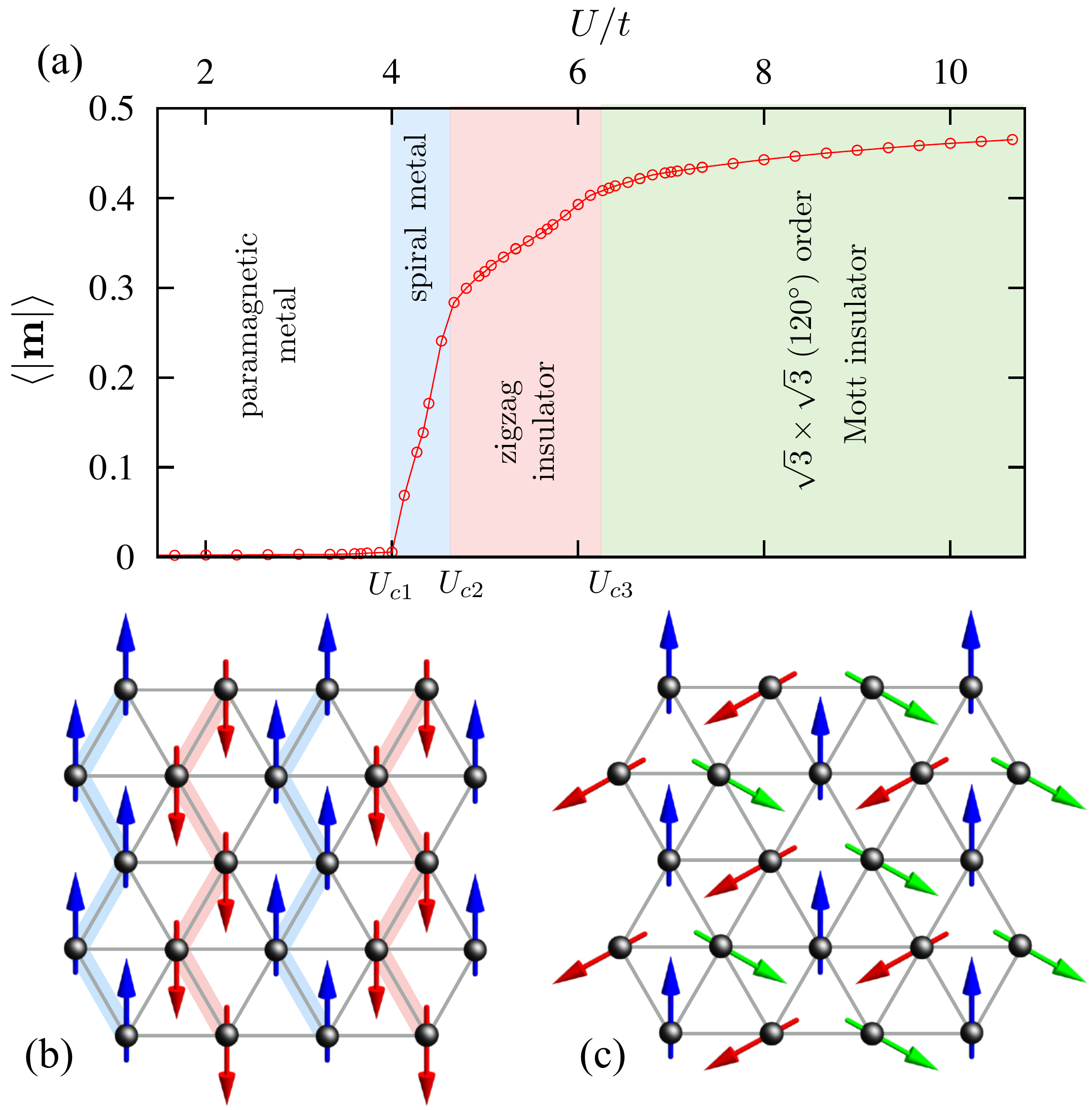}
\caption{(Color online) Phase diagram of the half-filled triangular-lattice Hubbard model with NN hopping $t$. With increasing Hubbard repulsion, the four different phases are: a paramagnetic metal, an incommensurate spiral metal, a collinear commensurate zigzag SDW, and the 120$^\circ$-order. (a) average spin length $\langle |\mathbf m| \rangle$ as a function of $U/t$. The two commensurate SDW phases, collinear zigzag and 120$^\circ$ orders, are shown in (b) and (c), respectively.
\label{fig:half-filling}}
\end{figure}

Instead of Markov-Chain Monte Carlo, here we employ the stochastic Ginzburg-Landau (GL) relaxation dynamics~\cite{hohenberg77,bray94} to sample the equilibrium SDW configurations within the static HS-field approximation,
\begin{eqnarray}
	\label{eq:dmdt_GL}
	\frac{d\mathbf m_i}{dt} = - \gamma \frac{\partial \langle \mathcal{H}_{\rm SDW} \rangle}{\partial \mathbf m_i }  + \bm \xi_i(t).
\end{eqnarray}
Here $\gamma$ is a damping constant, and $\bm\xi_i$ is a $\delta$-correlated fluctuating force satisfying $\langle {\bm \xi}_i(t) \rangle = 0$  and $\langle {\xi}_{i}^{\mu}(t) {\xi}_{j}^{\nu}(t') \rangle = 2 \gamma  k_B T \delta_{ij} \delta_{\mu\nu} \delta(t - t')$ for vector components $\mu$ and $\nu$. This equation is similar to the over-damped Langevin dynamics used in Ref.~\cite{barros13} for the Kondo-lattice model. We note that the magnitude $|\mathbf m_i|$ of the O(3) vector $\mathbf m_i$ is not fixed in this over-damped dynamics. A fictitious inertial mass term can be added to the above dynamics  to improve the efficiency of the simulation. Unlike conventional GL  simulations, where the force is given by the derivative of a phenomenological energy functional, here the force is computed by solving the equilibrium electron liquid of $\mathcal{H}_{\rm SDW}$ at each time-step. Or equivalently, the effective energy functional $\mathcal{E}_{\rm eff}(\{\mathbf m_i\}) = \langle \mathcal{H}_{\rm SDW} \rangle$ is obtained by integrating out electrons on the fly. Our method is thus similar to the so-called {\em quantum} molecular dynamics (MD) simulations, in which the inter-atomic force is computed by solving the quantum electron Hamiltonian~\cite{marx00}, instead of being derived from a phenomenological classical potential. Drawing on this analogy with MD simulations, our approach can be viewed as a quantum GL method for SDW. Interestingly, the quantum MD method in conjunction with the functional integral theory has already been employed to obtain complex magnetic orderings in itinerant magnet compounds in the past~\cite{kakehashi98}.

The GL method is particularly powerful when combined with our recently developed kernel polynomial method (KPM) and its gradient transformation in which forces acting on all spins can be efficiently computed~\cite{barros13}; see Ref.~\cite{supp1} for more details. The resulting linear-scaling KPM-GL method allows us to simulate large lattices with $N \simeq 10^5$ to $10^6$ sites. We apply the KPM-GL simulations to the triangular-lattice Hubbard model as a benchmark. The phase diagram shown in Fig.~\ref{fig:half-filling} agrees very well with those obtained by holon-doublon mean-field~\cite{krishnamurthy90} and the rotational-invariant slave-boson (SB)~\cite{capone01} calculations. Note that the SB method~\cite{kotliar86} describes both large and small $U$ regimes, and shows quantitative agreement with QMC  over a wide range of interaction and doping for the square-lattice Hubbard model~\cite{lilly90}.

At large $U$, our simulation finds the expected 120$^\circ$-order, which is the ground state of the Heisenberg Hamiltonian arising from the strong-coupling limit of the half-filled Hubbard model. In fact, the semi-classical Hamiltonian Eq.~(\ref{eq:H_SDW}) reduces to the {\it classical} Heisenberg spin model in the $U/t \gg 1$ limit. To see this, we note that any ground state for $t_{ij}=0$ contains exactly one electron in each site with its spin locally aligned with the SDW field $\mathbf m_i$. The magnitude of the SDW field freezes at $|\mathbf m_i|=1/2$ in this $U \to \infty$ limit because amplitude fluctuations have an energy cost proportional to $U$.  In analogy with the large $U$ limit of the original Hubbard model,  the ground state manifold is massively degenerate because each  $\mathbf m_i$ can point in an arbitrary direction. 
The degeneracy is removed to second order in $t_{ij}$. The low-energy effective Hamiltonian is obtained by considering virtual  electron hopping processes between two neighboring sites $i$ and $j$.  The Pauli exclusion principle dictates that an electron at site-$i$ can hop to the neighboring site only when its spin is {\em anti-aligned} with that of the local moment at site $j$. Consequently, the effective hopping constant between the two neighboring sites is $t^{\rm eff}_{ij} = t_{ij} \langle \chi_i | \chi_j \rangle = t_{ij} \sin(\theta_{ij}/2)$, where $|\chi_i\rangle$ are local electron spinor eigenstate and $\theta_{ij}$ is the angle between the two local moments. At second-order, the energy gain through the virtual electron hopping produces the effective interaction
\begin{eqnarray}
	\label{eq:H_ex}
	\mathcal{H}_{ij} = \frac{-2 t_{ij}^2 \sin^2(\theta_{ij}/2)}{U} = \frac{4 t_{ij}^2}{U} \left(\mathbf m_i \cdot \mathbf m_j - \frac{1}{4}\right).
\end{eqnarray}
This result corresponds to the classical limit of the $S=1/2$ Heisenberg model, implying that  the semi-classical SDW Hamiltonian (\ref{eq:H_SDW}) correctly captures the classical limit of the half-filled Hubbard model in the strong-coupling regime~\cite{fazekas99,supp2}.

As $U$ is decreased, our simulation shows that the 120$^\circ$ order is replaced by an interesting commensurate collinear SDW as the ground state~\cite{krishnamurthy90,capone01}. The collinear SDW at this intermediate $U$ is still gapped electronically and exhibits a zigzag structure. By computing the electron density of states (DOS), we find a metal-insulator transition at $U_{c2}$ between an incommensurate spiral and a commensurate collinear SDW phase (see Fig.~\ref{fig:half-filling}). Finally, the metallic spiral SDW undergoes a continuous transition at into a paramagnetic state at $U_{c1}$.
We note in passing that the metal-insulator transition, obtained with other numerical techniques (e.g. path-integral renormalization group method~\cite{yoshioka09}), is entirely within the paramagnetic regime~\cite{sahebsara08,yoshioka09}, implying the existence of a paramagnetic insulator (or spin liquid) at intermediate $U$ values. This state  cannot be obtained with our semi-classical approach because it is stabilized by strong fluctuations of the HS fields along the imaginary time axis. However, the existence of this phase remains to be settled. Recent variational QMC~\cite{tocchio13} and DMFT~\cite{goto16} calculations show that spiral SDW is more favorable than the spin liquid phase. Nonetheless, if a magnetic phase is stabilized by, e.g. applying a magnetic field, we expect the incommensurate and collinear SDWs obtained here to be strong candidates.

\section{Semiclassical dynamics and Landau-Lifshitz equation}
\label{sec:semiclassical}

Having demonstrated that the semiclassical Hamiltonian $\mathcal{H}_{\rm SDW}$ provides a viable approach to equilibrium SDW phases, a natural question is whether we can use it to study the  SDW dynamics. To this end, we need to reintroduce physical dynamics to the ``static'' auxiliary SDW field. We first note that the spin-fermion Hamiltonian $\mathcal{H}_{\rm SDW}$ can also be obtained from a Hartree-Fock decoupling of the interaction term that varies from one site to another. The important difference relative to the HS approach is that the SDW field satisfies the self-consistent condition $\mathbf m_i = \langle {\bf s}_i \rangle = {\rm Tr}(\rho \, \mathbf s_i)$, where $\rho$ is the density matrix characterizing the physical electron state. Consequently, the field $\mathbf m_i$ belongs to the sphere of radius $1/2$  ($0 \leq |\mathbf m_i| \leq 1/2$).

To derive the time dependence of the SDW field, we start with the continuity equation associated with the total spin conservation: $d \mathbf s_i /dt = - \sum_{j} \mathbf J_{ij}$, where $\mathbf s_i = \frac{1}{2} c^\dagger_{i\alpha} \bm\sigma_{\alpha\beta} c^{\;}_{i\beta}$ is the electron spin, and $\mathbf J_{ij} = -i\frac{t_{ij}}{2}  \bm\sigma_{\alpha\beta} (c^\dagger_{i\alpha} c^{\;}_{j\beta} - c^\dagger_{j\alpha}c^{\;}_{i\beta})$ is the spin current density on bond $\langle ij \rangle$. We then introduce the single-particle density matrix $\rho$ with elements $\rho_{i\alpha, j\beta} \equiv \langle c^{\dagger}_{j\beta}\,c^{\;}_{i\alpha} \rangle$. Taking the average of the continuity equation leads to 
\begin{eqnarray}
	\label{eq:dmdt}
	\frac{d\mathbf m_i}{dt} = -\frac{i}{2} \sum_j t_{ij} \bm\sigma^{\;}_{\beta\alpha} \left( \rho_{i\alpha, j\beta} - \rho_{j\alpha, i\beta} \right).
\end{eqnarray}
The SDW dynamics is thus related to the time evolution of the density matrix $\rho$, which obeys the von Neumann equation $d\rho/dt = i [\rho, H^{\rm eff}] $. Up to a constant, $H^{\rm eff}$ is the effective single-electron Hamiltonian defined as $\mathcal{H}_{\rm SDW} = \sum_{i\alpha, j\beta} H^{\rm eff}_{i\alpha, j\beta}\,c^\dagger_{i\alpha} c^{\;}_{j,\beta}$. Using Eq.~(\ref{eq:H_SDW}), we obtain the equation of motion  
\begin{eqnarray}
	\label{eq:drhodt}
	\frac{d\rho_{i\alpha, j\beta}}{dt} & =&  i \left(t_{ik} \, \rho_{k\alpha, j\beta} - \rho_{i\alpha, k\beta} \, t_{kj} \right)   \\
	 &+&  i U  \left(\mathbf m_i \cdot\bm\sigma_{\alpha\gamma}\, \rho_{i\gamma, j\beta} 
	 - \rho_{i\alpha, j\gamma} \, \bm\sigma_{\gamma\beta} \cdot \mathbf m_j \right). \nonumber
\end{eqnarray}
The electron density matrix is partly driven by the time-varying SDW field. Eqs.~(\ref{eq:dmdt}) and (\ref{eq:drhodt}) comprise  a complete set of coupled ordinary differential equations for the SDW dynamics. An alternative is to dispense of Eq.~(\ref{eq:dmdt}) and substitute $\mathbf m_i(t) =  \rho_{i\alpha, i\beta}(t) \bm\sigma_{\beta\alpha}$ in Eq.~(\ref{eq:drhodt}), giving rise to a set of nonlinear differential equations for $\rho$. 

In general, time dependence of physical quantities in mean-field approaches can be obtained using the Dirac-Frenkel variational principle~\cite{dirac30,frenkel34}. Our derivation here, on the other hand, is based on the spin-density continuity equation, hence emphasizing the importance of conservation laws in physical dynamics. This physically intuitive approach can be easily generalized to obtain dynamics for other symmetry-breaking phases.

Our approach here is a {\em real-space} formulation of the time-dependent Hartree-Fock (TDHF) method~\cite{mclachlan64,deumens94}, similar to the familiar time-dependent Bogoliubov-de~Gennes equation for superconductors or Bose condensates~\cite{degennes,volkov74}. Assuming that the order parameters are characterized by well defined momenta, e.g. $\mathbf m_i = \sum_r \mathbf M_r \, \exp(i \mathbf Q_r \cdot \mathbf r_i)$, the above equations can be simplified due to the translation invariance. The problem is then reduced to a set of coupled differential equations for density-matrix elements $n_{\alpha\beta}(\mathbf k, t) \equiv \langle c^\dagger_{\mathbf k\alpha}\,c^{\;}_{\mathbf k\beta} \rangle$ and $g^r_{\alpha\beta}(\mathbf k, t) \equiv (\mathbf M^{\;}_r \!\cdot \! \bm\sigma_{\alpha\beta}) \langle c^\dagger_{\mathbf k \alpha}   c^{\;}_{\mathbf k+\mathbf Q_r, \beta}\ \rangle $ in momentum space. This {\em mean-field} approximation of TDHF has recently been applied to the out-of-equilibrium dynamics of BCS-type superconductors~\cite{volkov74,barankov04}, and of N\'eel-type SDW~\cite{tsuji13,sandri13,blinov17}. 
Since the order-parameter field $\mathbf M_r$ is assumed to be uniform, spatial inhomogeneity and/or fluctuations are ignored in such $k$-space approaches. Our formulation here does not require the prerequisite knowledge of ordering patterns, and are particularly capable of simulating complex symmetry-breaking phases, inhomogeneous configurations, and disordered phases with  preformed local moments, such as the paramagnetic state in the large $U$ limit.

The SDW dynamics Eq.~(\ref{eq:dmdt}) can be simplified in the large $U$ limit in the so-called adiabatic approximation, which assumes that electrons quickly relax to the ground state of the instantaneous SDW configuration $\{\mathbf m_i\}$. Using second-order perturbation theory, one readily computes the average spin current: $ \langle \mathbf J_{ij} \rangle = \frac{4 t^2_{ij}}{U} \,\mathbf m_i \times \mathbf m_j$. Substituting this into the right-hand side of Eq.~(\ref{eq:dmdt}) gives rise to a Landau-Lifshitz (LL) equation 
\begin{eqnarray}
	\label{eq:dmdt_large_U}
	\frac{d\mathbf m_i}{dt} = -  \sum_j \frac{4 t_{ij}^2}{U}  \, \mathbf m_i \times \mathbf m_j,
\end{eqnarray}
with an effective torque computed using the Heisenberg exchange of Eq.~(\ref{eq:H_ex}). More details can be found in Ref.~\cite{supp2}.

For intermediate and small $U/t$ values, one needs to solve the von~Neumann equation. Since the number of independent density-matrix elements is of order $\mathcal{O}(N^2)$ for a lattice of $N$ spins, the computational cost of integrating the von Neumann equation is tremendous for large lattices, e.g. $N\sim 10^5$. To further simplify the numerical calculation, here we derive the SDW dynamics in a similar adiabatic limit for arbitrary $U$. Formally, we employ the multiple-time-scale method~\cite{davidson72} and introduce an adiabaticity parameter $\epsilon \sim |d\mathbf m/dt|$ such that the fast (electronic) and slow (SDW) times are $\tau = \epsilon t$ and $t$, respectively. The single-particle Hamiltonian varies with the slow time, i.e. $H^{\rm eff}(\{\mathbf m_i \}) = H^{\rm eff}(\tau)$. Expanding the density matrix in terms of the adiabaticity parameter: $\rho(t) = \rho^{(0)}(\tau) + \epsilon \rho^{(1)}(t, \tau) + \epsilon^2 \rho^{(2)}(t, \tau) + \cdots$, and plugging it into the von~Neumann equation, we obtain $[\rho^{(0)}, H^{\rm eff}] = 0$ and $\frac{d\rho^{(\ell)}}{dt} - i [\rho^{(\ell)}, H^{\rm eff}] = -\frac{d\rho^{(\ell-1)}}{d\tau}$ for $\ell \ge 1$. This provides a systematic approach to obtain the time dependence of the density matrix.

Here we use the leading adiabatic solution $\rho^{(0)}$ to compute the expectation value of the spin-current density $\mathbf J_{ij}$, which is the right-hand side of Eq.~(\ref{eq:dmdt}). We first write $H^{\rm eff} = T + \Sigma$ where $T_{i\alpha, j \beta} = - t_{ij} \delta_{\alpha\beta}$ is the tight-binding Hamiltonian and $\Sigma_{i\alpha, j \beta} = -U \delta_{ij} \, \mathbf m_i \cdot \bm \sigma_{\alpha\beta}$ is the spin-fermion coupling. It is then straightforward to show that Eq.~(\ref{eq:dmdt}) is simply $d {\mathbf m}_i/dt = i \bm\sigma_{\alpha\beta} [\rho, T]_{i \beta, i\alpha}/2$. Using the adiabatic equation $[\rho^{(0)}, H^{\rm eff}] = 0$, we have $[\rho^{(0)}, T] = -[\rho^{(0)}, \Sigma]$, which gives  
\begin{eqnarray}
	\frac{d\mathbf m_i}{dt} = -\frac{i U }{4} \bm\sigma_{\alpha\beta} \left[ \left(\bm\sigma^{\;}_{\beta\gamma} \rho^{(0)}_{i\gamma, i\alpha} - \rho^{(0)}_{i\beta,i\gamma} \bm\sigma^{\;}_{\gamma\alpha} \right)
	\cdot \mathbf m_i \right].
\end{eqnarray}
The right-hand side of the above equation can be further simplified using the properties of Pauli matrix multiplication: $\sigma^a \sigma^b = \delta_{ab} \mathcal{I}_{2\times 2} + i \epsilon_{abc} \sigma^c$, where $a, b, c$ are $x, y, z$. For example,  $\bm\sigma_{\alpha\beta}(\bm\sigma_{\beta\gamma} \cdot \mathbf m_i) \rho^{(0)}_{i\gamma,i\alpha} = n^{(0)}_i \mathbf m_i + i \mathbf m_i \times \bm\sigma^{\;}_{\alpha\beta} \rho^{(0)}_{i\beta, i\alpha}$, where $n^{(0)}_i = \rho^{(0)}_{i\alpha, i\alpha}$ is the local electron density. After some algebra, we obtain
\begin{eqnarray}
	\label{eq:dmdt_adiabatic}
	\frac{d\mathbf m_i}{dt} = \frac{U}{2} \mathbf m_i \times \bm\sigma^{\;}_{\beta\alpha}\, \rho^{(0)}_{i\alpha, i\beta}
	= - \mathbf m_i \times \frac{\partial \langle \mathcal{H}_{\rm SDW} \rangle}{\partial \mathbf m_i }.
\end{eqnarray}
The second equality comes from the fact that $\rho^{(0)}$ is computed from the equilibrium electron liquid described by $\mathcal{H}_{\rm SDW}$. The local  electron spin $\langle \mathbf s_i \rangle = \frac{1}{2} \langle c^{\dagger}_{i\alpha} \bm\sigma^{\;}_{\alpha\beta} \,c^{\;}_{i\beta} \rangle = \frac{1}{2} \bm\sigma^{\;}_{\beta\alpha}\, \rho^{(0)}_{i\alpha, i\beta}$ acts as an effective magnetic field and drives the slow dynamics of the SDW field.
Importantly, this equation shows that the adiabatic SDW dynamics is described by the Landau-Lifshitz (LL) equation~\cite{LL} with an effective energy functional $\mathcal{E}_{\rm eff}(\{\mathbf m_i \}) = \langle \mathcal{H}_{\rm SDW} \rangle$, obtained from the equilibrium electronic state of the instantaneous spin-fermion Hamiltonian.

\subsection{Benchmark with exact diagonalization}

\begin{figure}[t]
\centering
\includegraphics[width=0.88\columnwidth]{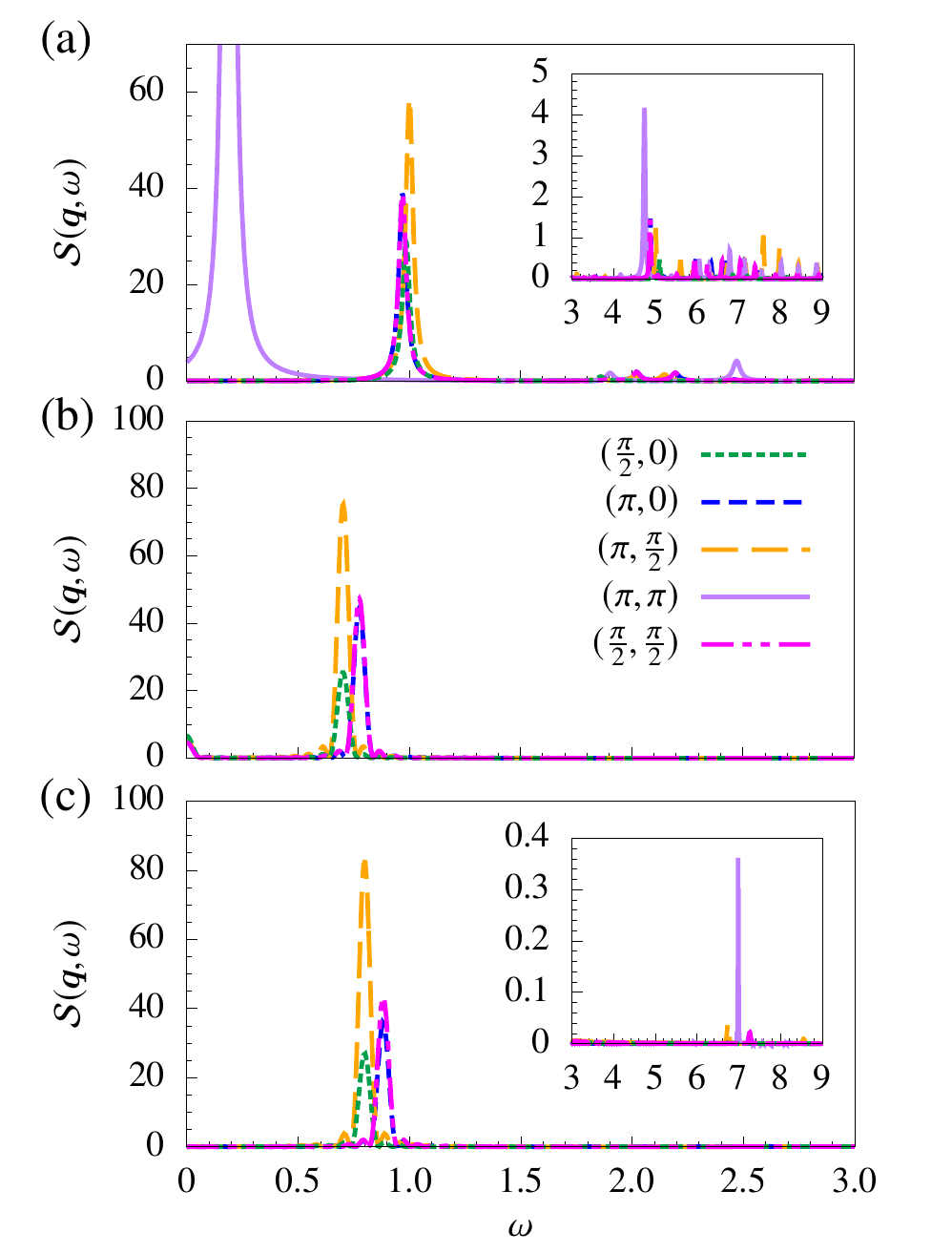}
\caption{ (Color online) Comparison of the dynamical structure factor of the Hubbard model on a $4\times4$ square lattice for $t=1$ and $U=7.33$: (a) The exact diagonalization result at $T=0$, where a Lorentzian broadening factor $\eta=0.02$ is used, (b) the semiclassical SDW calculation in the adiabatic approximation described by Eq.~(\ref{eq:dmdt_adiabatic}), and (c) the real-space TDHF calculation. For (b) and (c), the temperature was set $T=10^{-4}$ for generating initial configurations, and the results were averaged over 27 independent runs starting from different initial configurations.\label{fig:ED}}
\end{figure}

We first benchmark our semiclassical SDW dynamics, with and without the adiabatic approximation, against the exact diagonalization (ED) calculation of the original Hubbard model. To this end, we apply our formulation to the two-sublattice collinear N\'eel state that is obtained for the half-filled Hubbard model on a square lattice. Since we only include NN hopping, the N\'eel ordering is stable for any positive value of $U/|t|$.  Specifically, as shown in Fig.~\ref{fig:ED}, we compute the dynamical structure factor $\mathcal{S}(\mathbf k, \omega)$ of a $4 \times 4$ Hubbard cluster with periodic boundaries for  $U/t=7.33$. Details of the ED calculation are described in Ref.~\cite{supp3}.
We compare the ED result at $T=0$ and the semiclassical SDW dynamics at an extremely low temperature 
(classical moments freeze at $T=0$~\cite{note_A}). We  set temperature at $T=10^{-4}t$ and verify that  the results do not change upon decreasing the temperature to $T=10^{-5}$, indicating that our results  capture the dynamical response of the classical moments in the
$T \to 0$ limit.

For both the real-space TDHF dynamics [Eqs.~(\ref{eq:dmdt}) and (\ref{eq:drhodt})] and the adiabatic dynamics [Eq.~(\ref{eq:dmdt_adiabatic})], SDW states are first generated by means of GL-Langevin simulations described in Sec.~\ref{sec:GL-dyn}. The obtained spin configurations, which are representative of the canonical ensemble, are used as the initial condition for dynamical simulations. The dynamical structure factor, $\mathcal{S}(\mathbf k, \omega)$, is calculated by applying the space-time Fourier transform to the time evolution of the auxiliary field $\mathbf m_i(t)$. 
In the SDW dynamics,  the elastic peak has a finite width for finite duration of the dynamical simulation (not shown in Fig.~\ref{fig:ED}(b) and (c)). The area under the elastic peak is proportional to $N \langle {\mathbf m}_i \rangle^2$, where $N$ is the number of sites. In contrast, the lowest energy peak of the exact result appears at a small but finite frequency arising from quantum fluctuations neglected by the semiclassical treatment (the exact ground state is a singlet state for a finite size system). This quasi-elastic peak becomes the elastic peak of the spontaneously broken symmetry state in the thermodynamic limit. In Fig.~\ref{fig:ED}, we normalize the spectral weights obtained from the SDW dynamics so that the total weight of inelastic peaks equals that obtained from the ED excluding the quasi-elastic peak.

 The low-energy spectrum of the original Hubbard model is not well described by a simple effective spin Hamiltonian for $U/|t|=7.33$ (charge fluctuations can strongly renormalize the spin-wave dispersion). It is then quite remarkable that all the approaches produce a rather flat magnon dispersion for the wave-vectors included in a $4\times 4$ square lattice. However, the excitation energies in the adiabatic approximation are roughly 25\% lower than the exact result [see Fig.~\ref{fig:ED}]. This discrepancy is attributed to two factors: the semi-classical treatment of the spin degrees of freedom and the adiabatic approximation. The excitation energies obtained from the real-space TDHF method are  approximately 15\% lower than the exact result. We thus conclude that the adiabatic approximation accounts for roughly 10\% of the discrepancy, while the semiclassical  treatment accounts for the remaining 15\%. In addition, the normalized spectral weights (areas) in both the semiclassical dynamics are different only approximately 30\% from the exact result. A much better quantitative agreement is expected for 3D systems, but their solutions are beyond the scope of state of the art ED methods.

The real-space TDHF method captures not only the transverse modes, but also the longitudinal mode arising from {\it charge fluctuations}~\cite{chubukov92}. We note however that the method {\it does not capture} the longitudinal  spin fluctuations associated with {\it quantum fluctuations} of the magnetic moments. Unlike charge fluctuations, quantum (magnetic) fluctuations persist for arbitrarily large-$U/t$ (they arise from fluctuations of the ${\bf m}$-field along the imaginary time axis). These longitudinal fluctuations  correspond to  two-magnon excitations in a $1/S$ expansion~\cite{canali93}. Correspondingly, they have an energy of order $J \propto t^2/U$ for large $U/t$. In contrast, the longitudinal spin fluctuations arising from charge fluctuations lead  to the high-energy peaks at $\omega \sim U$, which are observed in the real-space TDHF dynamics, as shown in the inset of Fig.~\ref{fig:ED}(c), while they are absent in the adiabatic dynamics. Nevertheless, as expected for this value of $U/t$, the longitudinal mode is well separated from the transverse modes.

\subsection{120$^\circ$ SDW order in Hubbard and Anderson-Hubbard model}

Our benchmark study shows that both the TDHF and the adiabatic approach provide a reasonable description of the SDW dynamics. It is worth noting that the linearized TDHF equation of motion corresponds to the random phase approximation (RPA)~\cite{reinhard92,rowe70,pines88}. Our real-space formulation of the TDHF thus provides an efficient and universal numerical approach to describe nonlinear dynamics  beyond the RPA level. Moreover, our approach allows for computation of dynamical response functions  at any {\it finite temperature}, including the high-temperature regime in which the magnetic moments only exhibit short range correlations.

\begin{figure}[t]
\includegraphics[width=0.95\columnwidth]{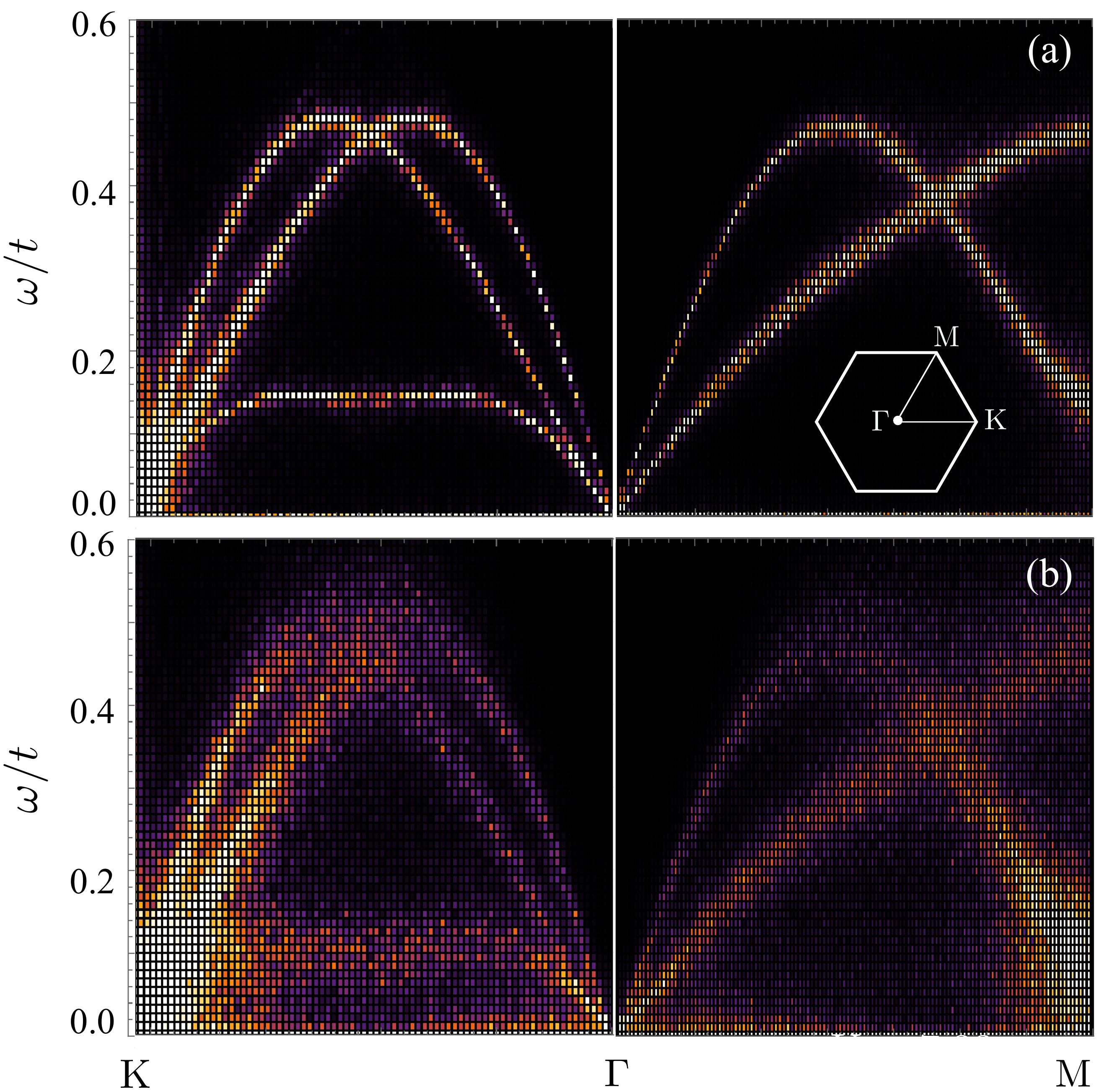}
\caption{(Color online) Dynamical structure factor $\mathcal{S}(\mathbf k, \omega)$ computed using the semiclassical SDW dynamics equation Eq.~(\ref{eq:dmdt_adiabatic}) for the 120$^{\circ}$-order in (a) Hubbard model and (b) Anderson-Hubbard model with an on-site disorder. The lattice size is $120 \times 120$. The  Hubbard parameter is $U = 7.33$ and the standard deviation of the on-site disorder is $\sigma_V = 0.45$, in units of NN hopping $t_{\rm nn}$.
\label{fig:struct_factor}}
\end{figure}

Another unique feature offered by our real-space method is the capability of computing the dynamical response of inhomogeneous SDW. To demonstrate this, here we apply the above adiabatic dynamics Eq.~(\ref{eq:dmdt_adiabatic}) to compute $\mathcal{S}(\mathbf k, \omega)$ for  the 120$^\circ$ SDW  depicted in Fig.~\ref{fig:half-filling}(c) for Hubbard model with quenched disorder.  Specifically, we consider the Anderson-Hubbard model by adding an on-site potential disorder $\sum_{i,\alpha} V^{\;}_i  c^\dagger_{i\alpha} c^{\;}_{i\alpha}$ to Eq.~(\ref{eq:H_SDW}). This model has served as a canonical platform for investigating the intriguing interplay of localization and correlations. Relevant to our study here is the effect of disorder on long-range SDW order. For N\'eel-type SDW on a half-filled bipartite lattice, it has been shown that increasing the disorder first closes the electron spectral gap, while the SDW remains finite~\cite{heidarian04}. The disappearance of the SDW order parameter occurs at a larger disorder~\cite{heidarian04,lahoud14}. This result is relevant to the non-equilibrium dynamics to be discussed below. 

We first compute the dynamical structure factor for the pure Hubbard model on a large lattice with $N = 120 \times 120$ sites. The results shown in Fig.~\ref{fig:struct_factor}(a) resemble the linear spin wave dispersion of the Heisenberg  model ~\cite{mourigal13}, except for a significantly renormalized lower-energy branch. 
Next we include a Gaussian disorder with zero mean and  standard deviation $\sigma_V = 0.45 \,t_{\rm nn}$, which is relatively large yet not strong enough to destroy the SDW order~\cite{byczuk09}.
The $\mathcal{S}(\mathbf k, \omega)$ computed using the adiabatic SDW dynamics is shown in Fig.~\ref{fig:struct_factor}(b). While the overall dispersion is similar to that of the SDW in the disorder-free Hubbard model [Fig.~\ref{fig:struct_factor}(a)], there are a few notable new features. Firstly, the magnon dispersion is significantly broadened by the quenched disorder, and the middle of the low-energy branch is further renormalized. Interestingly, a rather sharp dispersion remains near the zone center, indicating that these long-wavelength modes are less sensitive to disorder. Secondly, several new modes appear at low energies, especially below the original gap at the $M$ point. Interestingly, similar disorder-induced low-energy modes are also obtained in bi-layer Heisenberg antiferromagnet using the bond-operator method~\cite{vojta13}. 
A systematic study of the SDW dynamics with disorder will be left for future studies. 

\section{Nonequilibrium dynamics at finite temperatures}
\label{sec:noneq-SDW}

The adiabatic LL equation can also be used to study nonequilibrium SDW phenomena as long as the electron relaxation is much faster than the SDW dynamics. Here we first generalize the adiabatic dynamics to finite temperatures by adding dissipation and fluctuations to Eq.~(\ref{eq:dmdt_adiabatic}). It is worth noting that the adiabatic SDW dynamics preserves the length of local moments $|\mathbf m_i|$. Longitudinal spin relaxation and fluctuations thus come from either higher order terms in the adiabatic expansion or other processes beyond the self-consistent field approach. The standard Gilbert damping also preserves the spin length~\cite{gilbert55}. Instead, here we combine the Ginzburg-Landau relaxation discussed in Eq.~(\ref{eq:dmdt_GL}) with the adiabatic dynamics  of Eq.~(\ref{eq:dmdt_adiabatic}) to account for the longitudinal relaxation~\cite{solontsov93,hohenberg77}. This procedure gives rise to the following generalized LL equation
\begin{eqnarray}
	\label{eq:dmdt_SDW}
	\frac{d\mathbf m_i}{dt} = - \mathbf m_i \times \frac{\partial \langle \mathcal{H}_{\rm SDW} \rangle}{\partial \mathbf m_i } 
	- \gamma \frac{\partial \langle \mathcal{H}_{\rm SDW} \rangle}{\partial \mathbf m_i }  + \bm \xi_i(t).
\end{eqnarray}
 $\mathcal{H}_{\rm SDW}$ is the spin-fermion Hamiltonian defined in Eq.~(\ref{eq:H_SDW}), $\gamma$ is a damping constant, and $\bm\xi_i$ is a $\delta$-correlated fluctuating force satisfying $\langle {\bm \xi}_i(t) \rangle = 0$ and $\langle { \xi}^{\mu}_{i}(t) { \xi}^{\nu}_{j}(t') \rangle = 2 \gamma  k_B T \delta_{ij} \delta_{\mu\nu} \delta(t - t')$. The damping coefficient and the stochastic terms are chosen such that the dissipation-fluctuation theorem is satisfied and the above LL equation can be used to faithfully sample the equilibrium Boltzmann distribution at finite temperatures~\cite{ma12,kirilyuk10}. 

A microscopic calculation of the damping coefficient $\gamma$ is beyond the adiabatic approximation. In the real-space TDHF method, relaxation of SDW mainly arises from the Landau damping mechanism, which describes the energy transfer from the collective SDW mode to single-particle excitations~\cite{barankov06,blinov17}. Electron-electron scattering, which is not captured by the TDHF, also contributes to the damping of SDW, especially in ultrafast dynamics of metals~\cite{wingreen86,allen87}. Moreover, for open systems as in most pump-probe experiments, coupling of electrons to other degrees of freedom~\cite{suarez95,fatti00}, such as phonons, also play an important role in the relaxation of SDW dynamics. Here $\gamma$ is treated as as a phenomenological parameter which we chose to ensure the adiabatic approximation.  

That the SDW field obeys the LL dynamics can be understood intuitively from the fact that the Heisenberg equation of motion for spin operators corresponds to the classical LL equation~\cite{ma12}.  Here we give a microscopic derivation  starting from the Hubbard model, which reveals the condition for the validity of the LL dynamics (adiabatic approximation of the von Neumann equation). In fact,  the adiabatic approximation has been widely employed for spin dynamics in the context of time-dependent spin-density-functional theory~\cite{niu99,capelle01,qian02}. Our results thus provide a theoretical foundation for the LL dynamics of SDW, and pave the way for systematic improvements beyond the adiabatic approximation. 

It is worth noting that, in contrast to the conventional LL method, the energy functional in our approach is obtained by solving the spin-fermion Hamiltonian $\mathcal{H}_{\rm SDW}$ at {\em each} time-step. In analogy with the quantum MD simulations~\cite{marx00}, our numerical scheme can then be viewed as a quantum LL dynamics (QLLD) method. Although solving the electron Hamiltonian on the fly is computationally expensive, large-scale ($N \sim 10^5$) QLLD simulations are enabled by our recently developed KPM algorithm with automatic differentiation, such that the ``forces'' can be computed along with the total energy without extra overhead~\cite{barros13,supp1}.

We next apply our stochastic QLLD method to investigate the time evolution of a topological SDW on the triangular lattice that arises as a weak-coupling instability at filling fraction $n = 3/4$~\cite{martin08}. The combination of a van Hove singularity and perfect Fermi surface nesting at this filling fraction gives rise to a magnetic susceptibility that diverges as $\chi(\mathbf q) \propto \log^2| \mathbf q - \mathbf Q_\eta|$, where $\mathbf Q_\eta$ ($\eta = 1,2,3$) are the three nesting wavevectors~\cite{martin08}. The system thus tends to develop a triple-$\mathbf Q$ SDW characterized by three vector order parameters: $\mathbf m_i = \bm\Delta_1 e^{i\mathbf Q_1 \cdot \mathbf r_i} + \bm\Delta_2 e^{i\mathbf Q_2 \cdot \mathbf r_i} + \bm\Delta_3 e^{i\mathbf Q_3 \cdot \mathbf r_i}$. Note that the phase factors $e^{i \mathbf Q_\eta \cdot \mathbf r_i} = \pm 1$. In general, there are four distinct local moments, leading to a quadrupled magnetic unit cell.
The SDW instability of triangular-lattice Hubbard model at $n=3/4$ filling is similar to that of the half-filled Hubbard model on square lattice. However, unlike the simple N\'eel order in the later case, there are several possible triple-$\mathbf Q$ SDWs~\cite{barros13,nandkishore12}. 

\begin{figure}[t]
\includegraphics[width=0.9\columnwidth]{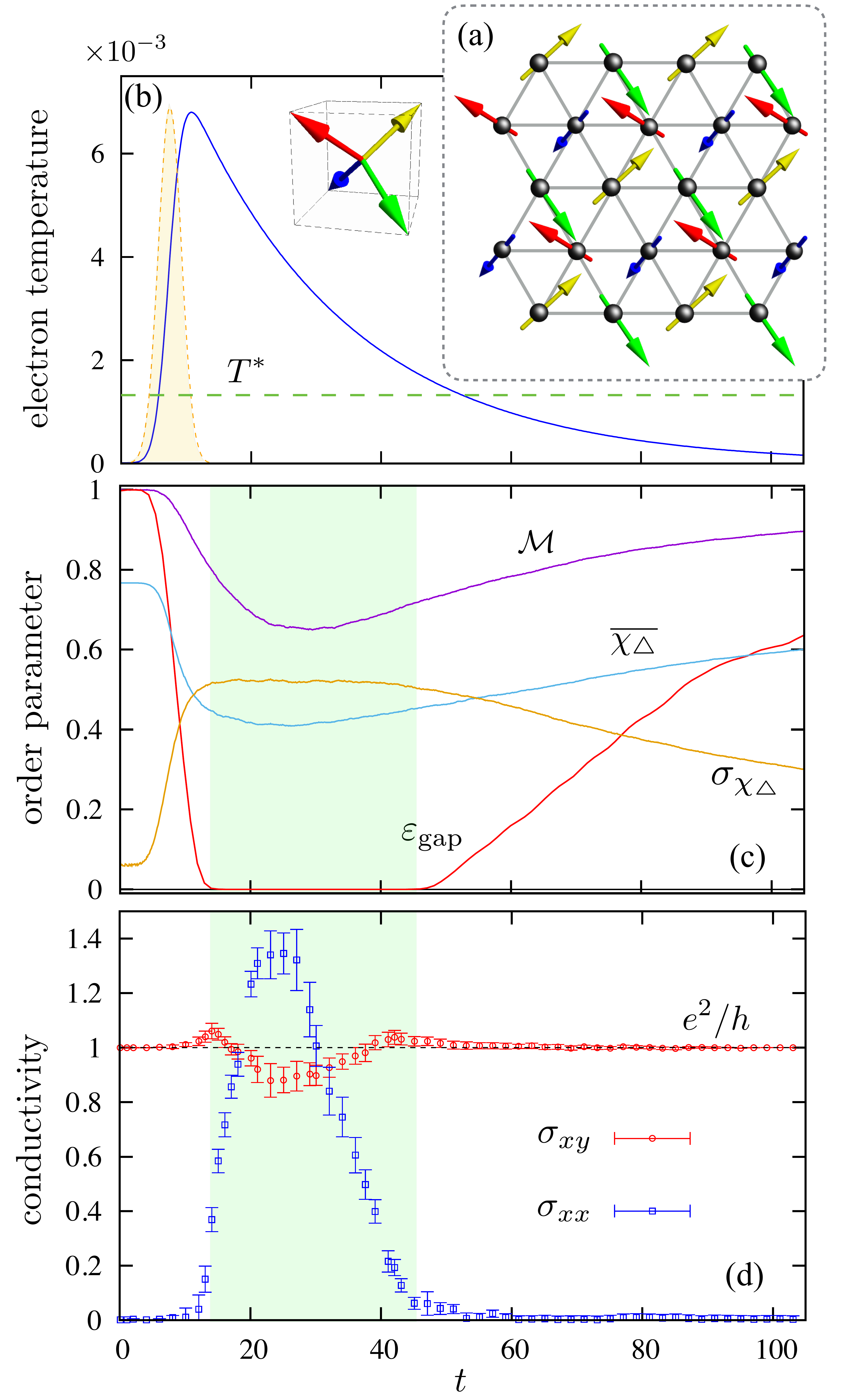}
\caption{(Color online) (a) Configuration of the topological triple-$\mathbf Q$ SDW with a quadrupled unit cell; the four spins in the unit cell point to corners of a regular tetrahedron. (b) Time dependence of the effective electron temperature. Also shown is the pump pulse with a Gaussian profile: $Q(t) \propto e^{-(t-t_p)^2/w^2}$, where $t_p = 15$, and $w = 5$, in units of $1/t_{\rm nn}$. The green dashed line marks the equilibrium transition temperature of the tetrahedral SDW. (c) the SDW order parameter $\mathcal{M}$, average normalized scalar spin chirality $\overline{ \chi_\triangle}$, standard deviation of the scalar chirality $\sigma_{ \chi_{\triangle}}$, and the electron energy gap $\varepsilon_{\rm gap}$ as a function of time. Both $\mathcal{M}$ and $\varepsilon_{\rm gap}$ are normalized to its maximum value. (d) the transverse and longitudinal conductivity (normalized to the quantized $e^2/h$) as a function of time. The (green) shaded area indicates the temporal window when the electronic gap is closed.
\label{fig:noneq_dyn}}
\end{figure}

At the lowest temperatures, the magnetic ordering consists of a non-coplanar SDW with $|{\bm \Delta}_1|=|{\bm \Delta}_2|=|{\bm \Delta}_3| $ and ${\bm \Delta}_1 \perp {\bm \Delta}_2 \perp {\bm \Delta}_3 $~\cite{martin08,nandkishore12,chern12}. This SDW is also called a tetrahedral or all-out order as spins in the unit cell point to the four corners of a regular tetrahedron~\cite{martin08}; see Fig.~\ref{fig:noneq_dyn}(a). Moreover, as the spins on each triangular plaquette are non-coplanar, the resulting nonzero scalar spin chirality $\mathbf m_i \cdot \mathbf m_j \times \mathbf m_k = \pm 4\Delta^3$ also breaks the parity symmetry. Consequently, the tetrahedral SDW is also characterized by a discrete $Z_2$ chirality order parameter. More importantly, electrons propagating in this non-coplanar SDW acquire a nonzero Berry phase, which is equivalent to a uniform magnetic field. Since the Fermi surface is gapped out by the SDW, the resulting electron state exhibits a spontaneous quantum Hall effect with transverse conductance $\sigma_{xy} = \pm e^2/h$~\cite{martin08,chern12}.

Motivated by  a recent pump-probe experiment on the ultrafast SDW dynamics in chromium~\cite{nicholson16}, we perform simulations of this topological SDW subject to a short heat pulse. For simplicity, we assume that the effect of the pump pulse is to inject energy to the electron system, which quickly equilibrates to a state characterized by temperature $T_e$. This is consistent with our adiabatic approximation for the SDW dynamics. The time dependence of the effective electron temperature is governed by the rate equation $C dT_e/dt = -G (T_e - T_L) + Q(t)$~\cite{anisimov74}, where $C$ is the heat-capacity of the electron liquid, $G$ is the coupling to the lattice, $T_L$ is the lattice temperature, and $Q(t) \propto \exp[-(t - t_p)^2 / w^2]$ is the heat source due to the pump pulse. We further assume that $T_L \approx 0$ throughout the relaxation process. The resultant $T_e(t)$ curve is shown in Fig.~\ref{fig:noneq_dyn}(b).

$T_e(t)$  is then used for the stochastic noise $\bm\xi(t)$ in our QLLD simulations of Eq.~(\ref{eq:dmdt_SDW}). We use the parameters, $U = 3$, damping $\gamma = 0.1$, $G/C = 0.02$, $t_p = 15$, and $w = 5$, in units of the NN hopping $t_{\rm nn}$. The lattice size is $N=120^2$. Fig.~\ref{fig:noneq_dyn}(c) shows the evolution of the magnetic order parameter at the nesting wavevectors: $\mathcal{M} = \sqrt{|\bm\Delta_1|^2 + |\bm\Delta_2|^2 + |\bm\Delta_3|^2}$ normalized to its maximum. We also estimate the time dependence of the electron spectral gap $\varepsilon_{\rm gap}$ from the instantaneous DOS [see Fig.~\ref{fig:noneq_dyn}(c)]. Interestingly, as the temperature rises, the decline of $\mathcal{M}$ is rather slow compared with the closing of the energy gap. In fact, the SDW order parameters $\bm\Delta_\eta$ remain finite throughout the process, while the gap closes quickly after the photoexcitation (at $t \approx 12$). In equilibrium the SDW order parameters disappear along with the gap above the transition temperature~\cite{note1}, implying that the  photoexcited SDW is in a highly non-equilibrium transient state. As the system relaxes, the gap reopens at a later time [see Fig.~\ref{fig:noneq_dyn}(c)]. 

This picture is further supported by our calculation of instantaneous longitudinal and transverse  conductivities  shown in Fig.~\ref{fig:noneq_dyn}(d). Here we use KPM to compute the Kubo-Bastin formula for  the conductivities~\cite{weisse06,garcia15}. The error bars  are estimated from five independent simulations. The electrons exhibit a negligible longitudinal conductivity $\sigma_{xx} \approx 0$ and a quantized Hall conductivity $\sigma_{xy} = e^2/h$ in the gapped regimes, as expected for this topological SDW. On the other hand, the longitudinal conductivity increases significantly during the period of vanishing gap, while the transverse conductance decreases  and exhibits small oscillations in the vicinity of the gap-closing transitions.

The closing and subsequent re-opening of the SDW gap  have been reported in recent pump-probe experiment on chromium~\cite{nicholson16}. The ultrafast SDW dynamics seem to be well described by a model that assumes  a thermalized electron gas. However, the closing of the gap is assumed to be always accompanied by the disappearance of the SDW order parameter in Ref.~\onlinecite{nicholson16}, which is not necessarily the case.  As demonstrated in our simulations, an out-of-equilibrium electron state might be gapless while the spin density remains modulated. Indeed, similar pump-probe experiments on the charge density wave (CDW) have revealed a fast collapsing of electronic gap in the time-resolved photoemission spectroscopy~\cite{perfetti06,petersen11}, and a reduced, yet finite, modulation of charge density inferred from core-level X-ray photoemission~\cite{hellmann10} during the nonequilibrium melting process. Numerical simulations taking into account coupling to the lattice distortion showed that the CDW order parameter can indeed be partially decoupled from the spectral gap dynamics~\cite{shen14}. However, it should be noted that the lattice degrees of freedom  introduce a new time scale, in addition to that of the hot electron relaxation. The transient metallic SDW observed in our simulations is probably due to a different mechanism.

To understand the origin of this nonequilibrium metallic SDW, we first note that the electronic gap of this topological SDW arises from the scalar spin chirality~\cite{chern12}. Indeed, the electronic gap vanishes for collinear or coplanar triple-$\mathbf Q$ SDWs~\cite{chern12,nandkishore12}. This observation leads us to investigate the temporal and spatial fluctuations of the scalar chirality. To this end, we introduce the normalized scalar spin chirality: $ \chi_{\triangle} =  \chi_{ijk} = (\mathbf m_i  \cdot  \mathbf m_j  \times  \mathbf m_k) / \, \overline{| \mathbf m |}^3$ for individual triangular plaquettes (here the overline indicates average over all triangles). The time dependence of the (spatial) average and the standard deviation of the scalar chirality $ \chi_{\triangle}$ are shown in Fig.~\ref{fig:noneq_dyn}(c). Interestingly, the average chirality remains finite and of the same sign, indicating that the chiral symmetry is still broken in this transient SDW. On the other hand, as shown in Fig.~\ref{fig:noneq_dyn}(c), the standard deviation $\sigma_{ \chi_\triangle}$ increases significantly with $T$.  In fact, the transient gapless regime coincides roughly with the period when $\sigma_{\chi_\triangle} > \overline{ \chi_\triangle}$, implying that the vanishing gap is due to thermally induced spatial fluctuations of $\chi_{ijk}$.

\begin{figure}[t]
\includegraphics[width=1\columnwidth]{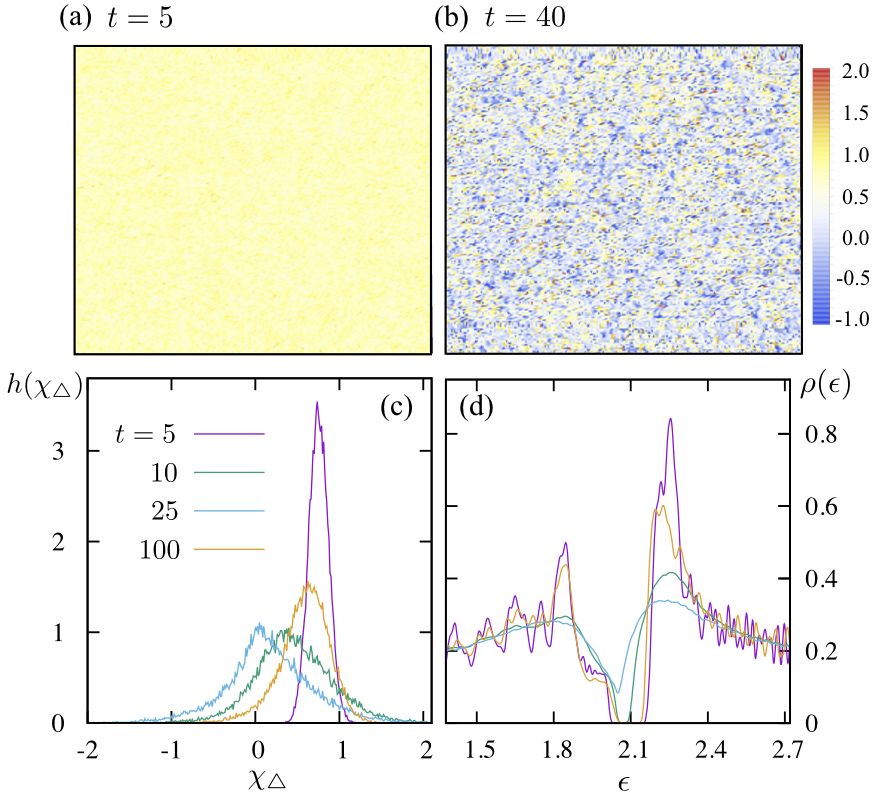}
\caption{(Color online) Top panels: spatial distribution of the normalized plaquette scalar spin chirality $ \chi_{\triangle}$ at time (a) $t = 5$ and (b) $t = 40$ from QLLD simulations on a triangular lattice with $120 \times 120$ sites. (c) histogram of the normalized plaquette chirality $ \chi_{\triangle}$ at varying simulation time. The corresponding electronic DOS near the Fermi level are shown in (d).
\label{fig:chirality}}
\end{figure}

Our scenario is confirmed by the spatial distribution of the normalized plaquette chirality at the initial stage ($t = 5$) and the gapless regime ($t  = 40$), shown in  Fig.~\ref{fig:chirality}(a) and (b), respectively. While the chirality is relatively uniform initially ($ \chi_\triangle \sim 1$), noticeable inhomogeneity develops at later times [see Fig.~\ref{fig:chirality}(b)]. Histograms of the plaquette chirality $h( \chi_\triangle)$ and the corresponding electron DOS at various simulation times are shown in Fig.~\ref{fig:chirality}(c) and (d), respectively. The chirality distribution becomes asymmetric and very broad during the period of vanishing gap. This transient gapless SDW is similar to the disorder-induced metallic antiferromagnetic state observed in the Anderson-Hubbard model~\cite{heidarian04}. As discussed above, the plaquette scalar chirality acts as local magnetic field and it is known that strongly disordered magnetic flux destroys the quantum Hall effect, in agreement with our simulations. Our results thus underscore the importance of thermal fluctuations and spatial inhomogeneity for the nonequilibrium dynamics of SDW, which have been overlooked in most dynamical studies of correlated systems.

\section{Summary and outlook}

We have developed a new theoretical framework for the semiclassical dynamics of SDW in Hubbard-like models. Based on a real-space time-dependent Hartree-Fock (TDHF) method applied to symmetry-breaking phases, our approach provides a Hamiltonian formulation for the SDW dynamics. The time evolution of the SDW field is coupled to the von~Neumann equation that describes the dynamics of single-electron density matrix. The formulation correctly reduces to the Holstein-Primarkoff dynamics of magnons (linear spin waves) in the large-$U$ limit at half-filling. We further show that an adiabatic approximation of the von~Neumann equation gives rise to a quantum Landau-Lifshitz dynamics (QLLD) for the SDW order parameter. Importantly, the energy functional of the LL equation is computed from an effective spin-fermion Hamiltonian that is obtained from a Hubbard-Stratonovich transformation of the original Hubbard model.

Our benchmark study of the N\'eel order on a half-filled Hubbard cluster showed that the semiclassical SDW dynamics agrees reasonably well with the exact diagonalization calculation. We apply our QLLD simulations to compute the dynamical structure factor of a 120$^\circ$ SDW at intermediate values of $U/t$ on the triangular lattice. While the overall spectrum resembles that obtained using linear spin-wave theory for the large $U/t$ limit of the Hubbard model ($S=1/2$ Heisenberg model),  charge fluctuations produce a significant renormalization of the low-energy branch. We note that quantum fluctuations, not included in our approach,  can also produce a significant renormalization of the spin-wave spectrum of frustrated 2D models \cite{chernyshev06,starykh06,zhitomirsky13,ma16}, like the one considered here. However, renormalization due to quantum fluctuations is much smaller in 3D models, whose dynamical structure factor is typically well described by semiclassical approaches.
Importantly, our real-space approach allows us to include the effects of spatial inhomogeneities of the SDW on large lattices. 
We have demonstrated this unique capability by computing the dynamical structure factor of the same 120$^\circ$ SDW on an Anderson-Hubbard model with disordered on-site potentials. Other than significant broadening of the magnon dispersion, our result shows that the disorder induces many low-energy modes, especially at the boundary of the Brillouin zone. 

Another important application of our QLLD method is the study of SDW-related non-equilibrium phenomena. Here we generalize the LL dynamics by including a Langevin-type damping and the corresponding stochastic noise to account for longitudinal relaxation and fluctuation. We then apply the generalized QLLD scheme to study the evolution of a topological SDW subject to a heat pulse, similar to the situation in the pump-probe setup. Our simulation shows an intriguing transient non-equilibrium SDW on the triangular lattice. While the SDW order parameter decreases with rising electron temperature, it remains finite even when the electronic gap is closed. The gap reopens at a later time as the system relaxes. Since the electronic gap in this topological SDW originates from the noncoplanar spin configuration, we show that the vanishing gap is due to strong spatial fluctuations of the scalar spin chirality, a quantity measuring the non-coplanarity of plaquette spins. Our real-space QLLD simulations thus underscores again the importance of spatial inhomogeneity and thermal fluctuation of the SDW dynamics.

The theoretical framework and numerical method developed in this paper can be easily generalized to study the dynamics of other symmetry-breaking phases, notably charge-density wave and superconductivity. Compared with other phenomenological method (e.g. time-dependent Ginzburg-Landau simulation for superconductors), keeping the electron degrees of freedom allows us to also look into the instantaneous electronic structure during the evolution of the order-parameter field. 

Our efficient semiclassical approach can also be feasibly integrated with first-principles method such as density functional theory (DFT). Here we note the analogy with molecular dynamics (MD) simulations. While classical MD simulations use phenomenological inter-atomic potentials, the quantum MD method computes the forces by solving the electron Hamiltonian on the fly. The quantum MD methods have proven a powerful tool in many branches of physical sciences. Our method can be viewed as the quantum version of the Landau-Lifshitz dynamics. We envision that our QLLD method combined with DFT calculation will provide a novel new approach to SDW dynamics in realistic materials.

It is worth pointing out that the numerical method presented here is complementary to DMFT. Both approaches are not restricted by the sign-problem that plagues the QMC methods. Conventional DMFT ignores spatial correlations from the outset and focuses on quantum effects or fluctuations along the imaginary time axis. Our semiclassical approach emphasizes the large-scale simulations in order to fully take into account the spatial  correlations and fluctuations of the magnetic order parameter. In developing this method, we are partly motivated by several recent studies emphasizing the important role of emergent nano-scale structures in the functionality of strongly correlated materials~\cite{dagotto05,fradkin15}.  Taking advantage of recent developments of efficient electronic structure method, such as KPM, our scheme is to progressively include the quantum corrections at each time step of the dynamical simulations. Our work here has laid the groundwork for systematic improvement beyond the adiabatic or TDHF approximation, which will be left for future studies.

\bigskip

\begin{acknowledgements}
The authors would like to thank A. Chubukov for useful discussions regarding analytical approaches to SDW dynamics. Work at LANL (K. B.) was carried out under the auspices of the U.S. DOE Contract No. DE-AC52-06NA25396 through the LDRD program. Part of the software and algorithm development  (G.W. Chern and C. D.  Batista) was supported by the Center for Materials Theory   as a part of the Computational Materials Science  (CMS) program, funded by the  U.S. Department of Energy, Office of Science, Basic Energy Sciences, Materials Sciences and Engineering Division.
\end{acknowledgements}

\appendix

\newpage
\pagebreak

\widetext
\begin{center}
\textbf{\large Supplemental Materials}
\end{center}

\section*{A. Kernel polynomial method and its gradient transformation}

The kernel polynomial method (KPM)~\cite{weisse06b} and the technique of automatic differentiation~\cite{barros13b} are crucial to our implementation of efficient QLLD simulations.  Here we briefly review these numerical techniques. Conventional KPM provides an efficient approach to computing the system free-energy $\mathcal{F}$. However, central to our QLLD simulations is the calculation of the `forces' acting on spins: $\partial \mathcal{F} / \partial \mathbf m_i$, where the effective energy functional is calculated from the quadratic fermion Hamiltonian: $\mathcal{F} = \langle \mathcal{H}_{\rm SDW} \rangle$. Specifically, the force is
\begin{eqnarray}
	\frac{\partial \mathcal{F} }{\partial \mathbf m_i} = -U \langle \mathbf s_i \rangle = -\frac{U}{2} \rho_{i\alpha, i\beta} \,\bm\sigma_{\beta\alpha}.
\end{eqnarray}
Computing the force is thus equivalent to evaluating the single-particle density matrix $\rho_{i\alpha, j\beta}$. We first introduce the single-particle Hamiltonian $H_{i\alpha, j\beta}$ such that the quadratic spin-fermion Hamiltonian is expressed as (up to a term that is independent of fermions)
\begin{eqnarray}
	\mathcal{H}_{\rm SDW} = \sum_{i\alpha,j \beta} H_{i\alpha,j\beta} c^{\dagger}_{i\alpha}\,c^{\;}_{j\beta} = \sum_{IJ} H_{IJ} c^{\dagger}_I \, c^{\;}_J.
\end{eqnarray}
Here we have introduced notation $I = (i, \alpha)$, $J = (j, \beta)$, $\cdots$ for simplicity. The density matrix is then given by the derivative
\begin{eqnarray}
	\rho_{IJ} = \langle c^{\dagger}_J \,c^{\;}_I \rangle = \frac{\partial \mathcal{F}}{\partial H_{IJ}}.
\end{eqnarray}
Next we outline the KPM procedure for computing the free energy which is expressed in terms of the DOS as $\mathcal{F} = \int \rho(\epsilon) \, f(\epsilon) d\epsilon$, where $f(\epsilon) = -T \log [1 + e^{-(\epsilon-\mu)/T}]$. KPM begins by approximating the DOS as a Chebyshev polynomial series,
\begin{eqnarray}
	\label{eq:rho_KPM}
	\rho(\epsilon) = \frac{1}{\pi \sqrt{1 - \epsilon^2}} \sum_{m=0}^{M-1} (2 - \delta_{0, m})  \mu_m\, T_m(\epsilon),
\end{eqnarray}
where $T_m(x)$ are Chebyshev polynomials, and $\mu_m$ are the expansion coefficients. The expansion is valid only when all eigenvalues of $H_{IJ}$ have magnitude less than one. This can in general be achieved through a simple shifting and rescaling of the Hamiltonian. Moreover, damping coefficients $g_m$ are often introduced to reduce the unwanted artificial Gibbs oscillations. Substituting $\rho(\epsilon)$ into the free energy expression gives
\begin{eqnarray}
	\label{eq:F_KPM}
	\mathcal{F} = \sum_{m=0}^{M-1} C_m\, \mu_m,
\end{eqnarray}
where coefficients $C_m = (2 - \delta_{0,m}) g_m \int_{-1}^{1} \frac{T_m(\epsilon) f(\epsilon)}{\pi \sqrt{1-\epsilon^2}} d\epsilon$ are independent of the Hamiltonian and may be efficiently evaluated using Chebyshev-Gauss quadrature. 

The key step of KPM is to replace computation of the Chebyshev moments $\mu_m = {\rm Tr}  \, T_m(H)$ by an ensemble average $\mu_m = \langle T_m(H) \rangle =  \frac{1}{R}\sum_{\ell=1}^{R} r_\ell^\dagger H r_\ell$ over random normalized column vectors $r$~\cite{silver94}. Taking advantage of the recursive relation of Chebyshev polynomials: $T_m(H) = 2 H \cdot T_{m-1}(H) - T_{m-2}(H)$, the moments can be evaluated recursively as follows:
\begin{eqnarray}
	\mu_m = r^\dagger \cdot \alpha_m,
\end{eqnarray}
where $r$ is a random vectors with complex elements drawn from the uniform distribution $|r_I|^2 = 1$. The random vectors $\alpha_m$ are given by
\begin{eqnarray}
	\label{eq:alpha}
	\alpha_m = \left\{\begin{array}{lcr} 
	r,  & \qquad & m = 0 \\
	H \cdot r, & \qquad & m=1 \\
	2 H \cdot \alpha_{m-1} - \alpha_{m-2}, & \qquad & m>1 \\
	\end{array}\right.
\end{eqnarray}
The above recursion relation also indicates that evaluation of $\mu_m$ that are required for computing $\mathcal{F}$ only involves matrix-vector products. For sparse matrix $H$ with $\mathcal{O}(N)$ elements, this requires only $\mathcal{O}(M N)$ operations, where $M$ is the number of Chebyshev polynomials. On the other hand, even with the efficient algorithm for $\mathcal{F}$, a naive calculation of the derivatives $\partial \mathcal{F}/\partial H_{IJ}$ based on finite difference approximation is not only inefficient but also inaccurate. The computational cost of finite difference is similar to the KPM-based Monte Carlo method with local updates.

To circumvent this difficulty, we employ the technique of automatic differentiation with reverse accumulation~\cite{griewank89}. Instead of directly using Eq.~(\ref{eq:F_KPM}), the trick is to view $\mathcal{F}$ as a function of vectors $\alpha_m$ and write
\begin{eqnarray}
	\label{eq:dFdH}
	\frac{\partial \mathcal{F}}{\partial H_{IJ}} = \sum_{m=0}^{M-1} \frac{\partial \mathcal{F}}{\partial \alpha_{m, K}} \frac{\partial \alpha_{m, K}}{\partial H_{IJ}},
\end{eqnarray}
Here $\alpha_{m, K}$ denotes the $K$-th component of vector $\alpha_m$, and summation over the repeated index $K$ is assumed. Using Eq.~(\ref{eq:alpha}), we have 
\begin{eqnarray}
	\label{eq:dalphadH}
	\frac{\partial \alpha_{0, K}}{\partial H_{IJ}} = 0, \qquad \frac{\partial \alpha_{1, K}}{\partial H_{IJ}} = \delta_{IK}\, \alpha_{0, J}, \qquad
	\frac{\partial \alpha_{m, K}}{\partial H_{IJ} } = 2 \delta_{IK}\, \alpha_{m-1, J}  \quad (m>1)
\end{eqnarray}
The expression of $\partial \mathcal{F}/\partial H_{IJ}$ can be simplified by introducing a new set of random vectors:
\begin{eqnarray}
	\beta_m \equiv \frac{\partial \mathcal{F}}{\partial \alpha_{m+1}},
\end{eqnarray}
From Eqs.~(\ref{eq:dFdH}) and~(\ref{eq:dalphadH}) , we obtain
\begin{eqnarray}
	\frac{\partial \mathcal{F}}{\partial H_{IJ}} = \beta_{0, I} \,\alpha_{0, J} + 2 \sum_{m=1}^{M-2} \beta_{m, I} \, \alpha_{m, J}.
\end{eqnarray}
Remarkably, the vectors $\beta_m$ can also be computed recursively. To this end, we note that the recursion relation~(\ref{eq:alpha}) implies that $\mathcal{F}$ depends on $\alpha_m$ through three paths:
\begin{eqnarray}
	\frac{\partial \mathcal{F}}{\partial \alpha_{m, K}} = \frac{\partial\mathcal{F}}{\partial \mu_m} \frac{\partial \mu_m}{\partial \alpha_{m, K}}
	+ \frac{\partial \mathcal{F}}{\partial \alpha_{m+1, L}} \frac{\partial \alpha_{m+1, L}}{\partial \alpha_{m, K}}
	+ \frac{\partial \mathcal{F}}{\partial \alpha_{m+2, L}} \frac{\partial \alpha_{m+2, L}}{\partial \alpha_{m, K}}.
\end{eqnarray}
The various terms above can be straightforwardly calculated:
\begin{eqnarray}
	\frac{\partial \mathcal{F}}{\partial \mu_m} = C_m, \quad \frac{\partial \mu_m}{\partial \alpha_{m, K}} = r^*_K, \quad
	\frac{\partial \alpha_{m+1, L}}{\partial \alpha_{m, K}} = 2 H_{LK}, \quad 
	\frac{\partial \alpha_{m+2, L}}{\partial \alpha_{m, K}} = -\delta_{LK}.
\end{eqnarray}
Consequently, 
\begin{eqnarray}
	\beta_m = C_{m+1} \, r^\dagger + 2 \beta_{m+1} \cdot H - \beta_{m+2}, \qquad (m < M-1).
\end{eqnarray}
Restoring the site and spin indices, we obtain the following expression for the density matrix
\begin{eqnarray}
	\rho_{i\alpha, j\beta} = \beta_{0, i\alpha} \,\alpha_{0, j\beta} + 2 \sum_{m=1}^{M-2} \beta_{m, i\alpha} \, \alpha_{m, j\beta}.
\end{eqnarray}
As in standard KPM, there are two independent sources of errors in our method~\cite{weisse06b,barros13b}: the truncation of the Chebyshev series at order $M-1$, and the stochastic estimation of the moments using finite number $R$ of random vectors.  The performance of the stochastic estimation can be further improved using correlated random vectors based on the probing method~\cite{tang12}. 
Most simulations discussed in the main text were done on a $120\times 120$ triangular lattice. The number of Chebyshev polynomials used in the simulations is in the range of $M = 1000$ to $2000$. The number of correlated random vectors used is $R = 64$ to $144$.

\section*{B. Large-$U$ limit: formal derivation} 

Here we present a formal derivation of the effective SDW Hamiltonian in the large-$U$ limit, which is expected to be equivalent to that of the original Hubbard model. Our first step is to write the spin-fermion Hamiltonian Eq.~(\ref{eq:H_SDW}) in a new reference frame, such that the local quantization axis of site $i$ coincides with the direction of the SDW field. Let $\mathbf m_i = |\mathbf m_i| (\sin\theta_i \cos\phi_i, \sin\theta_i \sin\phi_i, \cos\theta_i)$, the fermionic operators in the new reference frame are
\begin{eqnarray}
	\tilde c^\dagger_{i, +} &=& e^{-i \phi_i/2} \sin(\theta_i/2) \, c^\dagger_{i, \downarrow} + e^{i \phi_i /2} \cos(\theta_i/2)\, c^\dagger_{i, \uparrow},  \nonumber \\
	\tilde c^\dagger_{i, -} &=& e^{-i \phi_i/2} \cos(\theta_i/2) \, c^\dagger_{i, \downarrow} - e^{i \phi_i /2} \sin(\theta_i/2)\, c^\dagger_{i, \uparrow},
\end{eqnarray}
The inverse transformation is
\begin{eqnarray}
	c^\dagger_{i, \uparrow} = e^{-i \phi_i/2} \cos(\theta_i/2)\, \tilde c^\dagger_{i, +} - e^{-i \phi_i /2} \sin(\theta_i/2)\, \tilde c^\dagger_{i, -}, \nonumber \\
	c^\dagger_{i, \downarrow} = e^{+i \phi_i/2} \sin(\theta_i/2) \,\tilde c^\dagger_{i, +} + e^{+ i \phi_i/2} \cos(\theta_i/2) \, \tilde c^\dagger_{i, -}.
\end{eqnarray}
We separate the spin-fermion Hamiltonian into two parts $\mathcal{H}_{\rm SDW} = \mathcal{T} + \mathcal{U}$. The kinetic hopping term can be re-expressed in the new reference frame as 
\begin{eqnarray}
	\label{eq:T_expr}
	\mathcal{T} = - \sum_{\langle ij \rangle} \sum_{\mu\nu = \pm} t_{ij} \left(\eta^{\mu\nu}_{ij}\, \tilde c^\dagger_{i\mu} \, \tilde c^{\;}_{j \nu} + {\rm h.c.} \right),
\end{eqnarray}
where
\begin{eqnarray}
	\eta^{++}_{ij} &=& +\cos\left(\frac{\phi_j - \phi_i}{2}\right) \cos\left(\frac{\theta_j - \theta_i}{2} \right) + i \sin\left(\frac{\phi_j - \phi_i}{2} \right) \cos\left(\frac{\theta_j + \theta_i}{2} \right), \nonumber \\
	\eta^{+-}_{ij} &=& -\cos\left(\frac{\phi_j - \phi_i}{2}\right) \cos\left(\frac{\theta_j - \theta_i}{2} \right) - i \sin\left(\frac{\phi_j - \phi_i}{2} \right) \cos\left(\frac{\theta_j + \theta_i}{2} \right),  \nonumber \\
	\eta^{-+}_{ij} &=& -\cos\left(\frac{\phi_j - \phi_i}{2}\right) \cos\left(\frac{\theta_j - \theta_i}{2} \right) + i \sin\left(\frac{\phi_j - \phi_i}{2} \right) \cos\left(\frac{\theta_j + \theta_i}{2} \right),  \nonumber \\
	\eta^{--}_{ij} &=& +\cos\left(\frac{\phi_j - \phi_i}{2}\right) \cos\left(\frac{\theta_j - \theta_i}{2} \right) - i \sin\left(\frac{\phi_j - \phi_i}{2} \right) \cos\left(\frac{\theta_j + \theta_i}{2} \right).
\end{eqnarray}
To derive the effective Hamiltonian, we use the standard perturbation approach by treating the hopping $\mathcal{T}$ as a perturbation to the coupling term
\begin{eqnarray}
	\mathcal{U} = - U \sum_i |\mathbf m_i| \left( \tilde c^\dagger_{i,+} \tilde c^{\;}_{i, +} - \tilde c^\dagger_{i,-} \tilde c^{\;}_{i, -} \right) + U \sum_i |\mathbf m_i|^2.
\end{eqnarray}
For convenience, we first introduce the resolvent of  $\mathcal{U}$: $\hat G_0(\epsilon) = 1/(\epsilon - \mathcal{U} )$. The effective Hamiltonian up to second order in $t_{ij}$ is given by
\begin{eqnarray}
	 \mathcal{H}_{\rm eff} = \mathcal{P}\, \mathcal{T} \hat G_0(\epsilon_0) \mathcal{T} \mathcal{P},
\end{eqnarray}
where $\mathcal{P}$ is a projector onto the lowest energy subspace with one electron per site whose spin is parallel to local moment $\mathbf m_i$, and $\epsilon_0 = - NU/4$ is the energy of the degenerate large-$U$ ground state ($N$ is the number of lattice sites). Then, in the new reference frame, the only processes contributing to $\mathcal{H}_{\rm eff}$ are the spin-flip hoppings which annihilate electrons with spin $\mu=+$ and create electrons in a difference site with spin $\mu=-$. Given that each state of the lowest energy subspace is fully characterized by the field configuration $\{\mathbf m_i \}$, the effective Hamiltonian can be expressed in terms of the the SDW field:
\begin{eqnarray}
	\mathcal{H}_{\rm eff} &=& -\sum_{\langle ij \rangle} \frac{t_{ij}^2}{U} \left( \left|\eta^{-+}_{ij}\right|^2 + \left|\eta^{+-}_{ij} \right|^2 \right) \nonumber \\
	& = & \sum_{\langle ij \rangle} \frac{ t_{ij}^2}{U} \left[ 1 - \sin\theta_i \sin\theta_j \cos(\phi_i - \phi_j) - \cos\theta_i \cos\theta_j \right]  
	 =  \sum_{\langle ij \rangle} \frac{4 t_{ij}^2}{U} \left(\mathbf m_i \cdot \mathbf m_j - \frac{1}{4} \right)
\end{eqnarray}
which is the expected Heisenberg exchange interaction for spin-1/2 in the large $U$ limit~\cite{fazekas99b}.

We next derive the dynamics equation in the large-$U$ limit, which is given by the Landau-Lifshitz equation. To this end, we first consider the Heisenberg equation of motion for local spin operator $d \mathbf s_i / dt = -i [\mathbf s_i, \mathcal{H}_{\rm H}]$. Here $\mathbf s_i = \frac{1}{2} c^{\dagger}_{i\alpha} \bm\sigma^{\;}_{\alpha\beta} c^{\;}_{i\beta}$, and $\mathcal{H}_{\rm H}$ is the Hubbard Hamiltonian. Expressing the on-site $U$ interaction in terms of spin operators, it is given by\begin{eqnarray}
	\mathcal{H}_{\rm H} = - \sum_{\langle ij \rangle, \alpha} t_{ij} \left(c^\dagger_{i\alpha} c^{\;}_{j\alpha} + {\rm h.c.} \right) - \frac{2U}{3} \sum_i \mathbf s_i^2 + \frac{N_e U}{2}. 
\end{eqnarray}
where $N_e$ is the number of electrons. Obviously, the $U$ term of the Hubbard Hamiltonian commutes with the spin operator, and we have $d \mathbf s_i / dt =-i [\mathbf s_i, \mathcal{T}]$. For example, we consider the $z$ component first. Using commutation relations $[s^z_i, c^\dagger_{i, \uparrow}] = \frac{1}{2} c^{\dagger}_{i, \uparrow}$ and $[s^z_i, c^\dagger_{i, \downarrow} ] = -\frac{1}{2} c^\dagger_{i,\downarrow}$, we have
\begin{eqnarray}
	-i \left[s^z_i, \left(c^\dagger_{i \alpha} c^{\;}_{j \alpha} + c^\dagger_{j\alpha} c^{\;}_{i\alpha} \right) \right] 
	= -\frac{i \sigma_\alpha}{2} \left(c^\dagger_{i \alpha} c^{\;}_{j \alpha} - c^\dagger_{j\alpha} c^{\;}_{i\alpha} \right) 
	= -\frac{i}{2} \left(c^\dagger_{i \alpha} \hat\sigma^z_{\alpha\beta}\, c^{\;}_{j \beta} - c^\dagger_{j\alpha} \hat\sigma^z_{\alpha\beta}\, c^{\;}_{i\beta} \right).
\end{eqnarray}
Here $\sigma_\alpha = \pm 1$ for $\alpha = \uparrow, \downarrow$, respectively. Summing over repeated indices is also implied. The equation for the $x$ and $y$ components of $\mathbf s_i$ can be obtained by applying $\pi/2$ rotations to this equation. We have
\begin{eqnarray}
	\label{eq:dsdt}
	\frac{d\mathbf s_i}{dt} = - \sum_j \mathbf J_{ij},
\end{eqnarray}
where we have defined the spin current density operator $\mathbf J_{ij}$. 
\begin{eqnarray}
	\mathbf J_{ij} = -\frac{i \bm\sigma_{\alpha\beta}}{2}  \left(c^\dagger_{i\alpha} c^{\;}_{j\beta} - c^{\dagger}_{j\alpha} c^{\;}_{i\beta} \right).
\end{eqnarray}
It is worth noting that Eq.~(\ref{eq:dsdt}) is simply the continuity equation for the spin density. Taking the expectation value with respect to the ground state gives rise to the equation of motion for the SDW field
\begin{eqnarray}
	\label{eq:dmdt_avg_J}
	\frac{d\mathbf m_i}{dt} = -\sum_{j} \langle \mathbf J_{ij} \rangle,
\end{eqnarray}
Next we compute the expectation value of $\mathbf J_{ij}$ in the large $U$ limit. In the $t / U \to 0$ limit, obviously $\langle \mathbf J_{ij} \rangle = 0$. A nonzero contribution comes from the second-order perturbation due to electron hopping. The procedure is similar to what we did to derive the effective Hamiltonian. Specifically, we project the spin current operator into the degenerate low-energy manifold of $\mathcal{H}_{\rm SDW}$ with the electronic eigenstates corrected up to first order in the perturbation. For the $z$-component, we have
\begin{eqnarray}
	\langle J^z_{ij} \rangle = \frac{1}{2} \left[ \mathcal{P} J^z_{ij} G_0(\epsilon_0) \mathcal{T} \mathcal{P} + \mathcal{P}  \mathcal{T} G_0(\epsilon_0) J^z_{ij}  \mathcal{P} \right].
\end{eqnarray}
Once again, it is convenient to work in the new reference frame. For example, the current density operator
\begin{eqnarray}
	\label{eq:Jz_expr}
	J^z_{ij} = -\frac{i t_{ij}}{2} \sum_{\mu\nu = \pm} \left(\tau^{\mu\nu}_{ij} \tilde c^\dagger_{i \mu}\,\tilde c^{\;}_{j\nu} - (\tau^{\mu\nu}_{ij})^* \tilde c^\dagger_{j\nu} \,c^{\;}_{i\mu}\right).
\end{eqnarray}
Here we have introduced
\begin{eqnarray}
	\tau^{++}_{ij} &=& +\cos\left(\frac{\phi_j - \phi_i}{2}\right) \cos\left(\frac{\theta_j + \theta_i}{2} \right) + i \sin\left(\frac{\phi_j - \phi_i}{2} \right) \cos\left(\frac{\theta_j - \theta_i}{2} \right), \nonumber \\
	\tau^{+-}_{ij} &=& -\cos\left(\frac{\phi_j - \phi_i}{2}\right) \cos\left(\frac{\theta_j + \theta_i}{2} \right) - i \sin\left(\frac{\phi_j - \phi_i}{2} \right) \cos\left(\frac{\theta_j - \theta_i}{2} \right),  \nonumber \\
	\tau^{-+}_{ij} &=& -\cos\left(\frac{\phi_j - \phi_i}{2}\right) \cos\left(\frac{\theta_j + \theta_i}{2} \right) + i \sin\left(\frac{\phi_j - \phi_i}{2} \right) \cos\left(\frac{\theta_j - \theta_i}{2} \right),  \nonumber \\
	\tau^{--}_{ij} &=& +\cos\left(\frac{\phi_j - \phi_i}{2}\right) \cos\left(\frac{\theta_j + \theta_i}{2} \right) - i \sin\left(\frac{\phi_j - \phi_i}{2} \right) \cos\left(\frac{\theta_j - \theta_i}{2} \right).
\end{eqnarray}
Using expressions (\ref{eq:Jz_expr}) and similar one for the kinetic term Eq.~(\ref{eq:T_expr}), we obtain
\begin{eqnarray}
	\langle J^z_{ij} \rangle = -\frac{i t_{ij}^2}{U} \left\{ \left[ \tau^{-+}_{ij} (\eta^{-+}_{ij})^* - (\tau^{-+}_{ij})^* \eta^{-+}_{ij} \right]
	+ \left[ \tau^{+-}_{ij} (\eta^{+-}_{ij})^* - (\tau^{+-}_{ij})^* \eta^{+-}_{ij} \right]
	 \right\}
\end{eqnarray}
Using the definitions for $\tau^{\mu\nu}_{ij}$ and $\eta^{\mu\nu}_{ij}$, it can be shown that
\begin{eqnarray}
	\left[ \tau^{-+}_{ij} (\eta^{-+}_{ij})^* - (\tau^{-+}_{ij})^* \eta^{-+}_{ij} \right] &=& \left[ \tau^{+-}_{ij} (\eta^{+-}_{ij})^* - (\tau^{+-}_{ij})^* \eta^{+-}_{ij} \right] \nonumber \\
	&=& 2 i \sin(\phi_j - \phi_i) \left[ \sin^2 \left(\frac{\theta_j + \theta_i}{2} \right) - \sin^2 \left(\frac{\theta_j - \theta_i}{2} \right) \right] \nonumber \\
	&=& 2 i \sin(\phi_j - \phi_i) \sin\theta_i \sin\theta_j
\end{eqnarray}
In terms of the SDW field $\mathbf m_i$, the right-hand side of the above equation is $2 i \, \hat{\mathbf m}_i \times \hat{\mathbf m}_j \cdot \hat{\mathbf z}$, where $\hat{\mathbf m}_i$ is a unit vector along the local moment direction. Using the fact that $|\mathbf m_i| = 1/2$ in the large $U$ limit at half filling, this result indicates the following vector identity for the spin current
\begin{eqnarray}
	\langle \mathbf J_{ij} \rangle = \frac{4 t_{ij}^2}{U} \mathbf m_i \times \mathbf m_j.
\end{eqnarray}
Substituting this into Eq.~(\ref{eq:dmdt_avg_J}) gives the well known Landau-Lifshitz equation of motion (\ref{eq:dmdt_large_U}) in the main text for the Heisenberg exchange Hamiltonian.

\section*{C. Exact diagonalization calculation of dynamical structure factor}

Here we provide details of the exact diagonalization (ED) calculation of Hubbard Model on $4\times4$ square lattice with periodic boundary condition (PBC). For simplicity, we consider the case where $SU(2)$ symmetry is conserved in the model, which leads to $\mathcal{S}(\bm{q},\omega)=3\mathcal{S}^{zz}(\bm{q},\omega)$. 

To calculate $\mathcal{S}^{zz}(\bm{q},\omega)$ at $T=0$, we first obtain the ground state $|\Psi_{0}\rangle$ in the total $S_{z}=0$ sector at half-filling, by using the implicitly restarted Arnoldi method provided through the ARPACK libary \cite{lehoucq98}. The dynamical structure factor can be expressed through the fluctuation-dissipation theorem:
\begin{eqnarray}
S^{zz}(\bm{q},\omega) &=& -2 \text{Im} \chi_{zz}(\bm{q},\omega) \nonumber \\
&=& -2 \langle\Psi_{0}  | S_{\bm{q}}^z S_{-\bm{q}}^z | \Psi_{0} \rangle \, \text{Im} \langle \phi_0 | (\omega + i \eta + E_0-H)^{-1} | \phi_0 \rangle,
\end{eqnarray}
where $| \phi_0 \rangle \equiv S^{z}(-\bm{q})|\Psi_{0}\rangle / \sqrt{\langle\Psi_{0}  | S_{\bm{q}}^z S_{-\bm{q}}^z | \Psi_{0} \rangle}$, and $E_0$ is the ground state energy: $H |\Psi_0 \rangle = E_0 | \Psi_0 \rangle$.

The matrix inverse $(z-H)^{-1}$ can be calculated through the Lanczos algorithm~\cite{lanczos90,dagotto94}:
\IncMargin{1em}
\begin{algorithm}
	\SetKwData{Left}{left}\SetKwData{This}{this}\SetKwData{Up}{up}
	\SetKwFunction{Union}{Union}\SetKwFunction{FindCompress}{FindCompress}
	\SetKwInOut{Input}{input}\SetKwInOut{Output}{output}
	\Input{$|\phi_0 \rangle = S^{z}(-\bm{q})|\Psi_{0}\rangle / \sqrt{\langle\Psi_{0}  | S_{\bm{q}}^z S_{-\bm{q}}^z | \Psi_{0} \rangle}$, $b_0 = 0$}
	\BlankLine
	\For{$j = 0,1,2,\ldots$}{
		
		$\quad |w_j \rangle = H |\phi_j \rangle - b_j | \phi_{j-1} \rangle $;
		
		$\quad a_j = \langle w_j | \phi_j \rangle$;
		
		$\quad |w_j \rangle = |w_j \rangle - a_j | \phi_j \rangle$;
		
		$\quad b_{j+1} = \sqrt{\langle w_j | w_j \rangle}$;
		
		$\quad |\phi_{j+1} \rangle = |w_j \rangle / b_{j+1}$;
	}
	\caption{Lanczos algorithm}\label{Algo:lanczos}
\end{algorithm}\DecMargin{1em}

In the new basis $\{ |\phi_0\rangle, | \phi_1 \rangle, | \phi_2 \rangle,\ldots \}$, the Hamiltonian is expressed by a tridiagonal matrix:
\begin{equation}
H= 
\begin{pmatrix}
a_0 & b_1 & \\
b_1 & a_1  & b_2 \\
      & b_2 & a_2      & \ddots \\
      &        & \ddots & \ddots & b_n \\
      &        &            & b_n      & a_n 
\end{pmatrix}.
\end{equation}

With Cramer's rule, the first element of the inverse matrix can be expressed as a continued fraction:
\begin{equation}
\langle \phi_0 | (z-H)^{-1} | \phi_0 \rangle =
\left[ (z-a_{0})-\frac{b_{1}^{2}}{(z-a_{1})-\frac{b_{2}^{2}}{(z-a_{2})-\cdots}} \right]^{-1},
\end{equation}
which leads to
\begin{eqnarray}
	\mathcal{S}^{zz}(\bm{q},\omega) & =-2 \langle\Psi_{0}  | S_{\bm{q}}^z S_{-\bm{q}}^z | \Psi_{0} \rangle \cdot \text{Im} \left[ (z-a_{0})-\frac{b_{1}^{2}}{(z-a_{1})-\frac{b_{2}^{2}}{(z-a_{2})-\cdots}} \right]^{-1}, \label{eq:continued_fraction}
\end{eqnarray}
where $z\equiv\omega+i\eta+E_{0}$.


\begin{thebibliography}{99}
\bibitem{note1} Our equilibrium QLLD simulations find that both the triple-$\mathbf Q$ SDW order parameters and the energy gap vanish above a transition temperature $T^* \approx 0.0015$.
\end{thebibliography}
\end{document}